# TRACE ELEMENT DISTRIBUTION IN HUMAN TEETH BY X-RAY FLUORESCENCE SPECTROMETRY AND MULTIVARIATE STATISTICAL ANALYSIS

CRISTIANA OPREA[1], PAVEL J. SZALANSKI[2], MARINA V. GUSTOVA[3], IOAN A. OPREA[1], VIOLETA L. BUZGUTA[4]

[1]*Frank Laboratory of Neutron Physics, Joint Institute for Nuclear Research, 141980 Dubna, RO*

[2]*Faculty of Physics and Applied Informatics, Lodz University, Lodz, Poland*

[3]*Flerov Laboratory of Nuclear Reactions, Joint Institute for Nuclear Research, 141980 Dubna, RF*

[4]*Dental Clinic "Artisan", 3700 Oradea, Romania*

*Abstract.* X-ray fluorescence spectrometry (XRFS) was used as a multielement method of evaluation of individual whole human tooth or tooth tissues for their amounts of trace elements. Measurements were carried out on human enamel, dentine, and dental cementum, and some differences in tooth matrix composition were noted. In addition, the elemental concentrations determined in teeth from subjects of different ages, nutritional states, professions and gender, living under various environmental conditions and dietary habits, were included in a comparison by multivariate statistical analysis (MVSA) methods. By factor analysis it was established that inorganic components of human teeth varied consistently with their source in the tissue, with more in such tissue from females than in that from males, and more in tooth incisor than in tooth molar.

## 1. INTRODUCTION

The municipal town of Oradea, located in the north-western part of Romania, on the Crisul Repede river bank and about 18 kilometres east of the frontier with Hungaria, is a medium industrial centre crossed by the traditional European highways. The environmental situation of this urban area, including the exposure to the ambient emissions of its inhabitants, is enough complexly since the accelerated development of the town throughout the industrial advancement. The urban activity in Oradea is undergoing transition since the closure of some industrial plants in the late 1990s and due to the all processes of modernization connected with the adhering of Romania to the European Union.

Usually the concentrations of trace heavy metals in teeth are higher in urban areas compared with those found in the teeth of people living in rural areas. As toxic heavy metals are easily deposited in tooth tissues by replacing the mineral tooth compounds during the human life, they can be used as indicators of cumulative long-term contamination with heavy metals of the studied population [1,2].

The first author started some monitoring environmental studies of this area together with a group of researchers of the Joint Institute for Nuclear Research and affiliated research institutions under the agreement of National Agency for Scientific Research (Romania) in 2003. During this period, we collected data on various environmental and ecological parameters in this region. Then we investigated the application of nuclear and atomic methods to trace element analysis of human teeth from Oradea's population, and published several papers on on the obtained results. The heavy metal content in deciduous teeth of the population segment under study was found to refer to some environmental parameters indicating recent industrial pollution [3-5].

The present work reflects the elemental concentrations of human teeth and tooth tissues measured by XRFS method and the most significant features underlined by multivariate analysis of the experimental data.

## 2. MATERIALS AND METHODS

### 2.1. Experimental material

The experimental material used in this study was prepared based on decayed permanent teeth from urban inhabitants of Oradea and from 5 of the residents leaving in a remote area located at 60 km north of Oradea, in a similar manner with the work [6]. The teeth from several nonsmoking donors of different age, gender and profession residing in a given area all their life were collected. The patient history included information about habit, medication, possible chronic diseases and others. The teeth sampling included the all tooth types recorded from males as well as from female subjects. The decayed teeth chosen for research are represented only by their own physical and chemical characteristics as they were not subjected to any stomatological treatment in the past. The analytical samples included the all tooth (n=116) as well as thick layers of tooth enamel (n=95), dentine (n=55) and cementum (n=22). The general characteristics of the teeth survey from which the data were obtained are shown in Table 1.

Table 1.
Characteristics of the human teeth survey

| Tooth tissue | Group | Age range | Tooth type | | | | |
|---|---|---|---|---|---|---|---|
| | | | Incisor | Canine | Premolar | Molar | Wisdom tooth |
| **All tooth** | Humans | 18-72 | 23 | 13 | 29 | 33 | 18 |
| **Enamel** | Female | 18-68 | 13 | 3 | 12 | 19 | 9 |
| | Male | 18-72 | 3 | 1 | 13 | 17 | 5 |
| **Dentine** | Female | 18-55 | 2 | 2 | 7 | 10 | 1 |
| | Male | 18-65 | 3 | 4 | 15 | 11 | |
| **Cementum** | Female | 18-52 | 1 | 1 | 3 | 4 | - |
| | Male | 18-58 | 1 | 1 | 4 | 5 | 2 |

## 2.2. Analytical procedure

The teeth were cleaned by any extraneous material (i.e. tartar, caries and soft tissue) and further rinsed with distilled water. Then they were dried until constant weight in a thermostat at $105^0C$ and stored in individual polyethylene containers. Experimental samples of dental enamel, dentine and cementum tissues were cut in layers of 1-2 mm thickness and used further in the analytical procedures.

The whole tooth and tooth tissue measurements were made by using a XRFS set-up which was proven to provide concentration data which were suitable enough for statistical analysis [7]. Dental standard materials of $Ca_{10}(PO_4)_6(OH)_2$ and $Ca(OH)_2$ were measured in the same experimental conditions as the dental tissue samples and Ca from the both materials was chosen as a standard for validation of the analytical results. Results of the analyses for all element concentrations are expressed in µg/g of dry tooth weight.

## 2.3. MVSA

The Ca-corrected concentrations for the all tooth and teeth layers were 10log transformed and subjected to multivariate analysis to test for differences between human subjects in a similar manner as in the work [5]. The analytical data were subjected to several procedures of multivariate analysis to search the dependence between different characteristics of human life, as gender and age of the urban residents and also tooth tissues and types, and elemental concentrations.

The statistical methods used for multivariate modeling of concentration data were R-mode correlation coefficients [8], varimax method, Kaiser normalization and exploratory factor analysis. The calculations were performed by using Microsoft Excel program and SPSS program package for Windows.

## 3. RESULTS AND DISCUSSION

The results of element concentrations measured (whole tooth measurements) in the teeth of urban population segment were normalized to Ca and then to the element concentrations of the control group (Figure 1). The concentrations for most of the metals were significantly higher for the Oradea inhabitants compared with those of rural group. The highest ratios were found for Zn (6.65), Fe (5.68), In(Cd) (4.75), Nd (3.83), As (3.67) and Cr (3.36). Therefore, the question may arise whether differences in the tooth metal concentrations is related to the individual's urban exposure to possible sources of trace heavy metals [4,5].

The most significant element concentrations determined in tooth tissues for the tooth from which we were able to separate the three slices, namely enamel, dentine and cementum, also reflected the changes in tooth matrix composition (Fig. 2).

It is well-known that the mineral component of enamel is 95%, of dentine is 67% and that of cementum is about 45-50% and the rest of each toot tissue includes organic material and water [9-11]. This supports the tendency reflected by the determined concentrations, that ($conc_{enamel} > conc_{dentine} > conc_{cementum}$) for almost all inorganic elements. Exception from this behavior is done by Zn and K, for which ($conc_{dentine} > conc_{enamel}$).

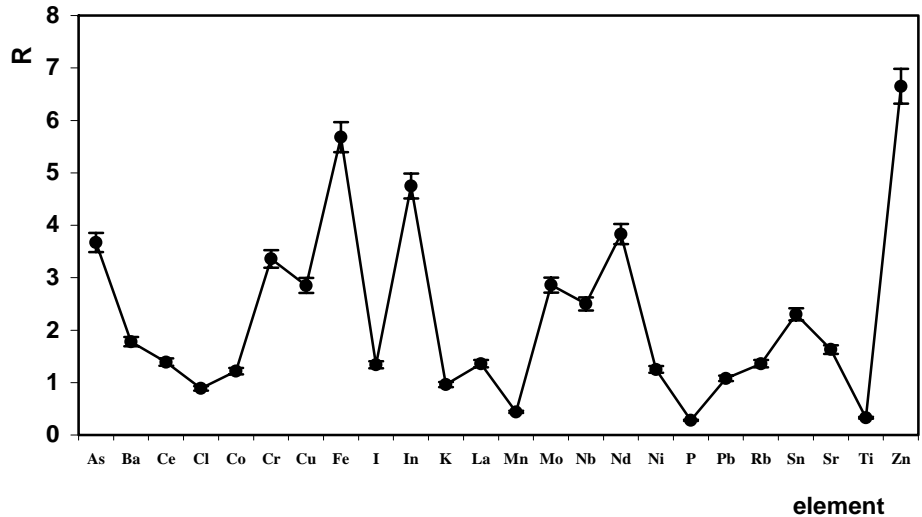

Figure 1. Average element concentrations in deciduous teeth of urban population versus those of the control population

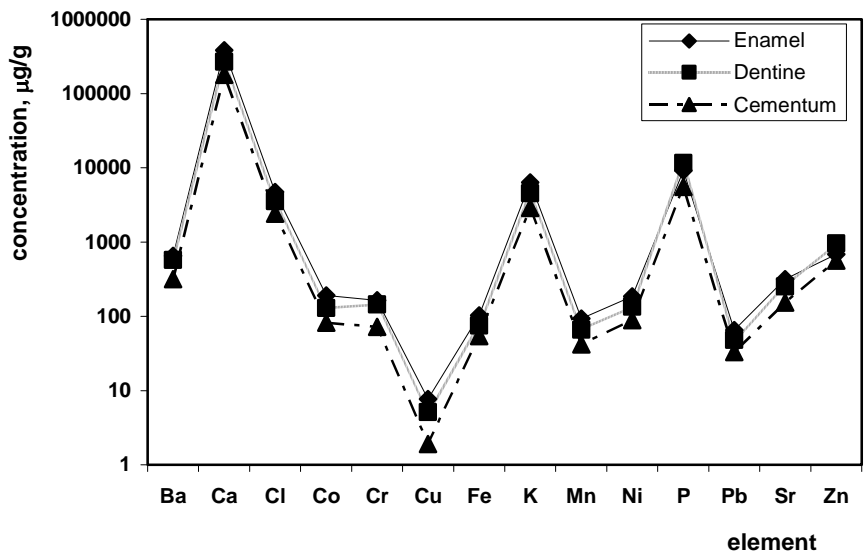

Figure 2. The mean concentrations of the most significant inorganic constituents in tooth tissues

The Oradea residents of various ages were assigned the following four groups: 18-25, 25-35, 35-60 and >60 years old. Only data from these age groups were statistically analyzed, because they were known to represent different age periods of the human life, respectively, for each individual and thus could be compared across age groups and between subjects. These ranks were further used as coded values for the variable "age" in multivariate statistical evaluations.

The Ca-corrected concentrations of 10 metals were regressed against age, gender, tooth type and toot tissue by exploratory statistical analysis to reveal the population related trends in trace metal data. Product-moment R-mode correlation matrix based on the interrelationships between the analyzed variables is listed in Table 2. The correlation matrix shows significant positive correlations of Fe, Cr, Cu, Zn and Ni, but a negative correlation of Fe and Cr with the "age", tooth type and tooth tissue.

Table 2.
Product-moment R-mode correlation matrix of the concentrations of significant elements, age, gender, tooth type and toot tissue in deciduous teeth of Oradea population segment

| Variable | Fe | Cr | Cu | Zn | As | Mn | Ni | V | Co | Sr | Age | Gender | Tooth type |
|---|---|---|---|---|---|---|---|---|---|---|---|---|---|
| Fe | 1.00 | | | | | | | | | | | | |
| Cr | 0.82 | 1.00 | | | | | | | | | | | |
| Cu | 0.46 | 0.42 | 1.00 | | | | | | | | | | |
| Zn | 0.40 | 0.29 | 0.71 | 1.00 | | | | | | | | | |
| As | 0.33 | 0.13 | 0.42 | 0.46 | 1.00 | | | | | | | | |
| Mn | 0.54 | -0.18 | 0.48 | 0.32 | -0.14 | 1.00 | | | | | | | |
| Ni | 0.52 | 0.38 | 0.53 | 0.41 | 0.29 | 0.07 | 1.00 | | | | | | |
| V | 0.28 | 0.20 | 0.24 | 0.36 | 0.35 | 0.22 | 0.41 | 1.00 | | | | | |
| Co | 0.08 | 0.13 | 0.32 | 0.29 | 0.23 | -0.10 | 0.13 | 0.15 | 1.00 | | | | |
| Sr | -0.05 | -0.05 | 0.35 | 0.24 | 0.21 | -0.07 | 0.28 | -0.18 | 0.10 | 1.00 | | | |
| Age | -0.40 | -0.32 | 0.30 | -0.16 | 0.38 | 0.52 | 0.34 | 0.24 | 0.14 | 0.43 | 1.00 | | |
| Gender | 0.26 | 0.14 | 0.21 | 0.35 | 0.30 | 0.21 | 0.19 | -0.05 | -0.06 | 0.09 | 0.16 | 1.00 | |
| Tooth type | -0.15 | -0.09 | -0.05 | -0.14 | 0.09 | 0.07 | 0.13 | -0.02 | -0.01 | -0.16 | 0.03 | 0.64 | 1.00 |
| Toot tissue | -0.32 | -0.17 | -0.13 | 0.37 | -0.16 | -0.13 | 0.08 | 0.13 | 0.03 | -0.22 | 0.19 | 0.28 | 0.85 |

Significance: $p \leq 0.05$

To gain insight into the interrelations of the variables measured an R-mode exploratory factor analysis on the data set was applied. On the ground of multivariate procedure and Kaiser Criterion (eigenvalue>1) three relevant factors were retained accounting for almost 75% of the total variation. The application of varimax rotation of the standardized component loadings leads to improving of the separation in factor loadings variance (Figs. 1-3). Closely related variables Fe, Cr, Cu, V and Ni clustered near Factor 1 ("Power industry" factor) together with the tooth tissue. F1 accounted for 44.7% of the total variance in the data set. Despite large differences in the metal concentrations, the population segment presented increased fractions of Fe, Cr and Cu. Such trace metals are dietary and essential to proper human body functioning. However, high levels may substantially accumulate in hard tissues, including teeth, and result in toxic effects. This interpretation is supported by the specific emissions from power plants stations of CET 1 and CET 2 in the last 3 - 4 decades, as shown in the works [4,5] and certified by the earlier published documents [12,13].

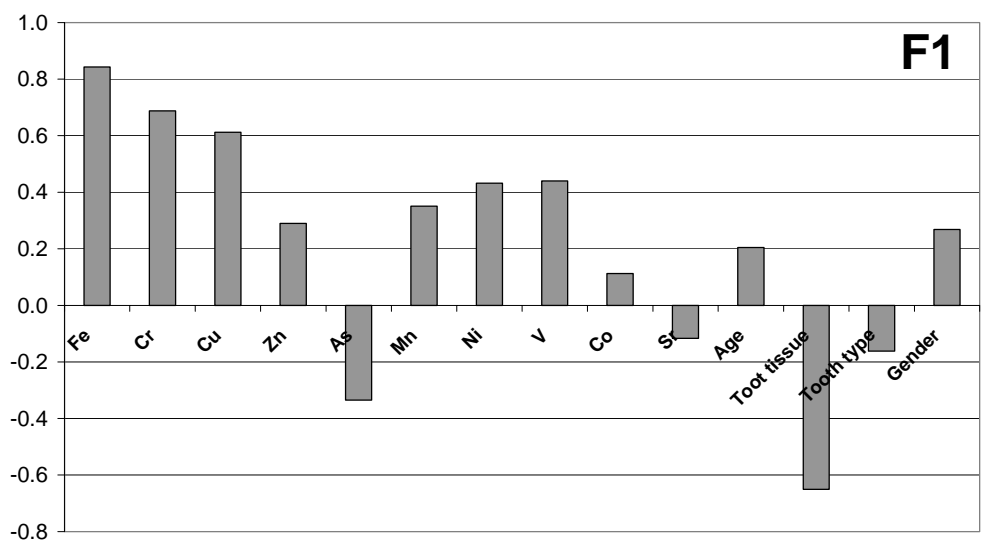

Figure 3. Exploratory factor F1 determined by R-mode MVSA

The second factor was characterized by large fractions of Cu, Zn, As, Mn and tooth tissue and accounted for 17.3% from the total variance of the data set. Apart of dietary uptake, the transition metals occurrence closely relates to an overall contamination signal due to the environmental or occupational exposure.

The human characteristics as age, tooth tissue, tooth type and gender centered around factor 3. It accounted for 13% of the total variance. The factor is loaded also by Co and dietary Sr, which are slightly correlated with tooth characteristics. This fact may be connected with the interactions of mineral compounds of teeth with bivalent metals, resulting by mineralization of the teeth in the polluted habits (Wakamura et al., 1998).

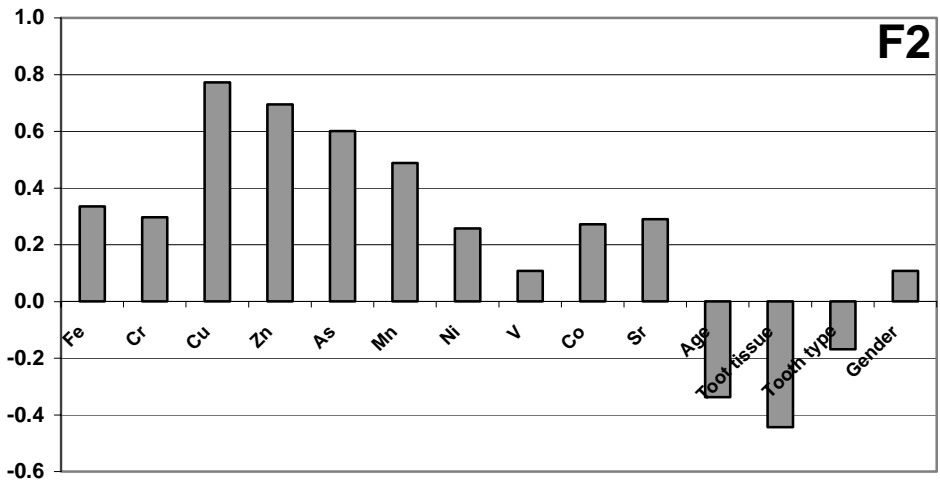

Figure 4. Exploratory factor F2 determined by R-mode MVSA

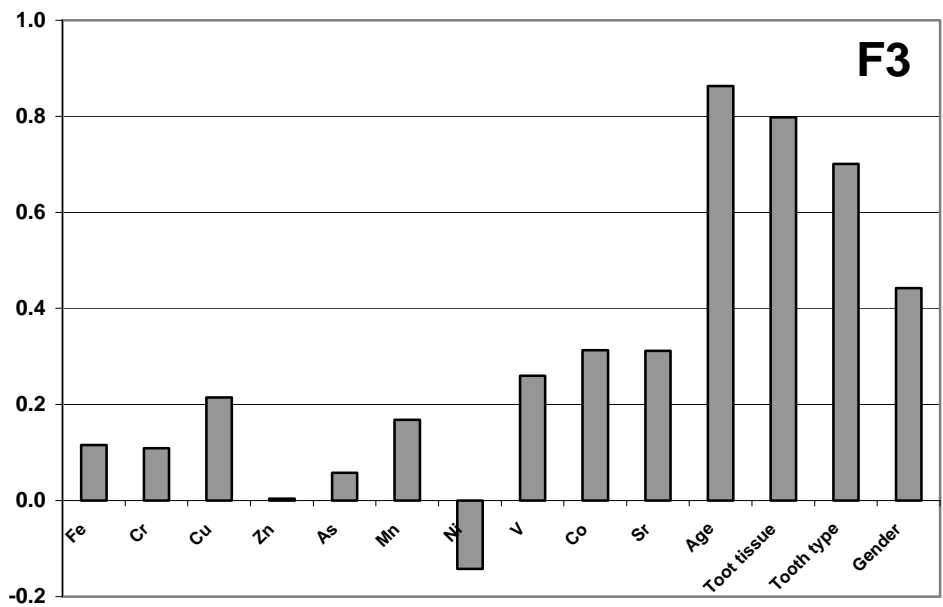

Figure 5. Exploratory factor F3 determined by R-mode MVSA

By exploratory factor analysis it was established that the inorganic components of human teeth varied consistently with their source in the tissue, with more in such tissue from females than in that from males, and more in incisor than in molar tissues. The heavy metal concentrations in the tissue did not varied greatly from one dental tissue to the other, whatever the age and gender of the subject.

## 4. CONCLUSIONS

The diversified environmental exposure of the investigated human segment was reflected by the concentrations of the trace metals in teeth. The results demonstrated that XRFS supplemented by MVSA is a useful and practical approach for the investigation of trace heavy metal incorporation and distribution on the surface of teeth as well as in inner layers. The multivariate statistical analyses performed seem to indicate that deciduous teeth might be a suitable indicator of environmental exposure to several trace elements.

The Product-moment R-mode correlation matrix and exploratory Factor analysis extracted three factors describing the concentration data set. The first factor, in order of its importance, was characterised by pollutant elements coming from the power industry as incorporated in the tooth tissues; the second factor was loaded with transition metals incorporated as well in the toot tissues and differentiated by age. The third factor has a human nature.